\documentclass[conference]{IEEEtran}

\usepackage[mathcal]{eucal}
\usepackage{mathrsfs}

\usepackage{cite}

\usepackage[pdftex]{graphicx}
\usepackage[cmex10]{amsmath}
\usepackage{amssymb}
  \usepackage[caption=false,font=footnotesize]{subfig}

\usepackage{bm}
\usepackage{amssymb}

\usepackage{epstopdf}
\newcommand{\argmin}{\mathop{\rm arg~min}\limits}

\usepackage{color}

\def\BibTeX{{\rm B\kern-.05em{\sc i\kern-.025em b}\kern-.08em
    T\kern-.1667em\lower.7ex\hbox{E}\kern-.125emX}}
\begin{document}

\title{Wireless 3D Point Cloud Delivery\\ Using Deep Graph Neural Networks}

\author{
\IEEEauthorblockN{
Takuya Fujihashi\IEEEauthorrefmark{1},
Toshiaki Koike-Akino\IEEEauthorrefmark{2},
Siheng Chen\IEEEauthorrefmark{2},
Takashi Watanabe\IEEEauthorrefmark{1}
}

\IEEEauthorblockA{
\IEEEauthorrefmark{1}Graduate School of Information Science and Technology, Osaka University, Japan\\
}
\IEEEauthorblockA{
\IEEEauthorrefmark{2}Mitsubishi Electric Research Laboratories (MERL), 201 Broadway, Cambridge, MA 02139, USA
}
}

\maketitle

\begin{abstract}
    In typical point cloud delivery, a sender uses octree-based digital video compression to send three-dimensional (3D) points and color attributes over band-limited links. 
    However, the digital-based schemes have an issue called the cliff effect, where the 3D reconstruction quality will be a step function in terms of wireless channel quality.
    To prevent the cliff effect subject to channel quality fluctuation, we have proposed soft point cloud delivery called HoloCast. Although the HoloCast realizes graceful quality improvement according to wireless channel quality, it requires large communication overheads. 
    In this paper, we propose a novel scheme for soft point cloud delivery to simultaneously realize better quality and lower communication overheads. The proposed scheme introduces an end-to-end deep learning framework based on graph neural network (GNN) to reconstruct high-quality point clouds from its distorted observation under wireless fading channels.
    We demonstrate that the proposed GNN-based scheme can reconstruct clean 3D point cloud with low overheads by removing fading and noise effects. 
\end{abstract}


\section{Introduction}
Three-dimensional (3D) holographic displays~\cite{bib:holodisplay,bib:holodisplay2} have emerged as attractive interface techniques for reconstructing 3D perceptual scenes that provide full parallax and depth information for human eyes.
3D holographic display can be widely used for many applications: entertainment, virtual training, and medical imaging.
Specifically, such 3D holographic visualizations will play a more important role in the post-Coronavirus (COVID-19) society because the 3D data can realize high-presence in remote conferencing~\cite{bib:zooms} and healthcare~\cite{bib:6G}. 
For example, holographic data of doctors and medical imaging provide more interactive verbal guidance in tele-surgery~\cite{bib:telesurgery}.

Point cloud~\cite{bib:mpeg} is one of data formats to represent 3D scenes/objects on the holographic display~\cite{bib:hologram}. 
We focus on wireless point cloud delivery systems, which send 3D point data to a remote display over wireless links to reproduce the corresponding 3D scenes/objects.
In contrast to conventional two-dimensional (2D) images, 3D points in point cloud data are massive, non-ordered, and non-uniformly distributed in space.
One of major issues in point cloud delivery is how to compress and send such numerous and irregular structure of 3D points while keeping original 3D scenes/objects. 
For example, when the number of 3D points is $800{,}000$, the amount of traffic without any compression will be approximately $38$~Mbits~\cite{bib:datasize}. 
Large traffic causes low 3D reconstruction quality in point cloud delivery over limited data rate links in wireless communications. 

\begin{figure}[t]
  \begin{minipage}{0.5\textwidth}
  \centering
   \subfloat[Holographic Display~\cite{bib:holodisplay3}]{\includegraphics[width=0.7\linewidth, trim=0 100 0 0, clip]{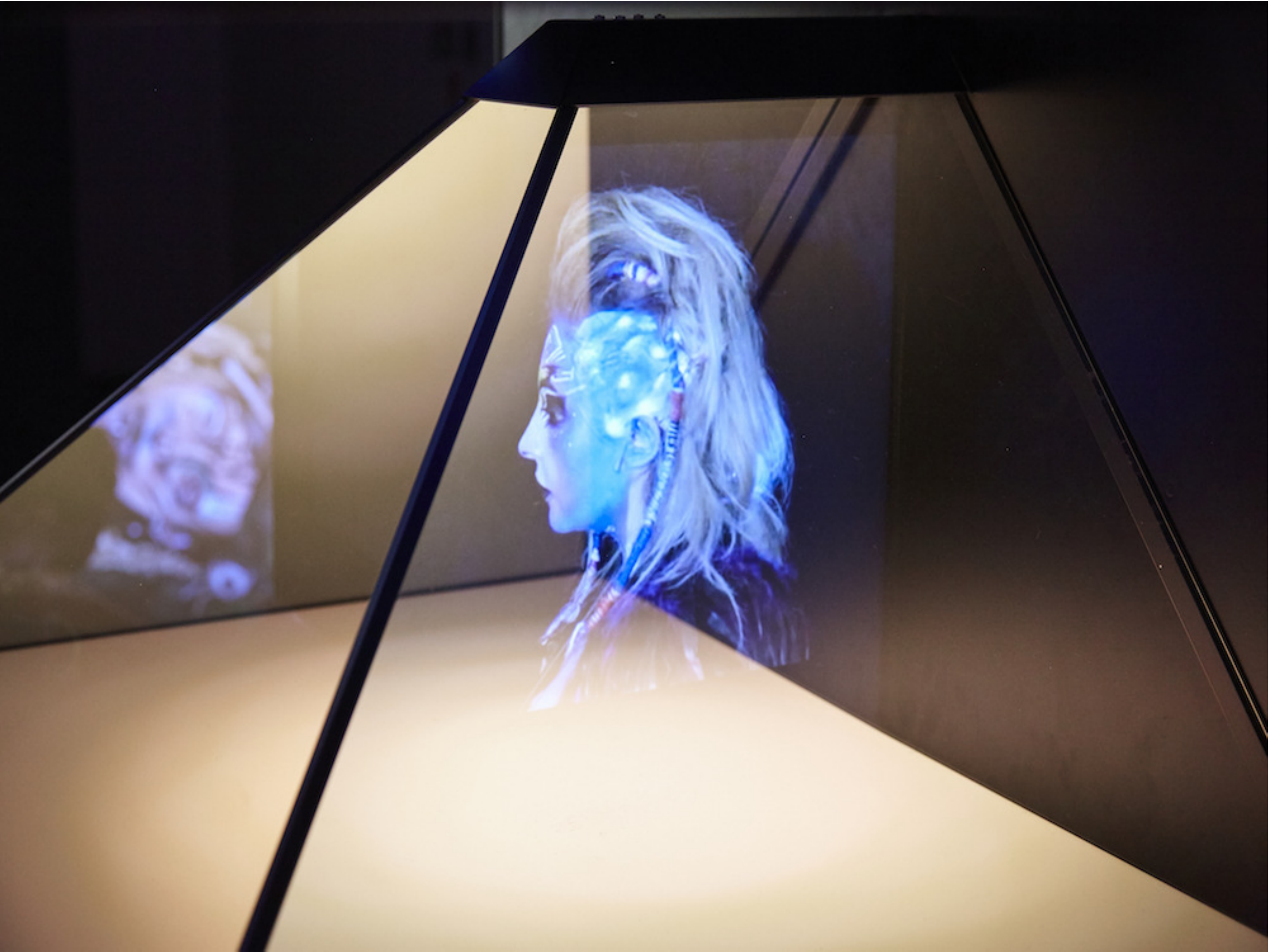}} \\
   \subfloat[3D Modeling~\cite{bib:3DScene}]{\includegraphics[width=0.7\linewidth]{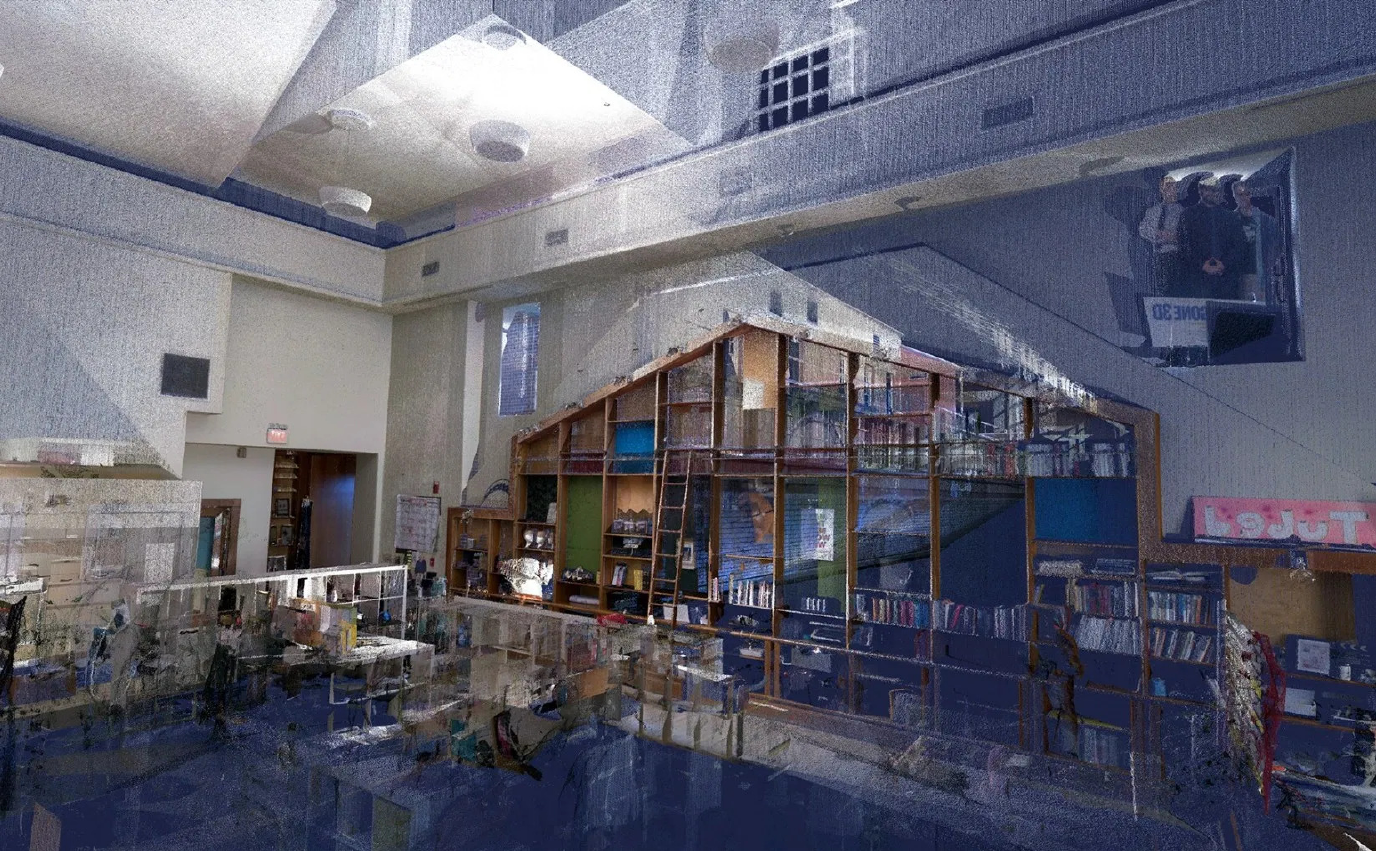}}
\end{minipage} \hfill
 \caption[]{Examples of 3D point applications.}
 \label{fig:vrar}
\end{figure}

For point cloud compression over wireless links, conventional schemes typically rely on digital encoding such as point cloud library (PCL)~\cite{bib:PCL, bib:PCL2}. 
Specifically, a sender decomposes point cloud into multiple 3D point sets by using the octree decomposition~\cite{bib:octree}, quantizes, and takes entropy coding to generate the compressed bitstream. 
Here, the compression rate of the bitstream is adaptively selected according to link capacity of wireless channels.
The compressed bitstream is then transmitted over the channels by using a channel coding and digital modulation scheme.
A successful high-quality delivery of point clouds over wireless links can realize high presence in video applications such as virtual reality and augmented reality on wireless devices as shown in Fig.~\ref{fig:vrar}.

\begin{figure*}[t]
  \centering
  \subfloat[Conventional GFT-Based HoloCast]
  {\includegraphics[width=\linewidth]{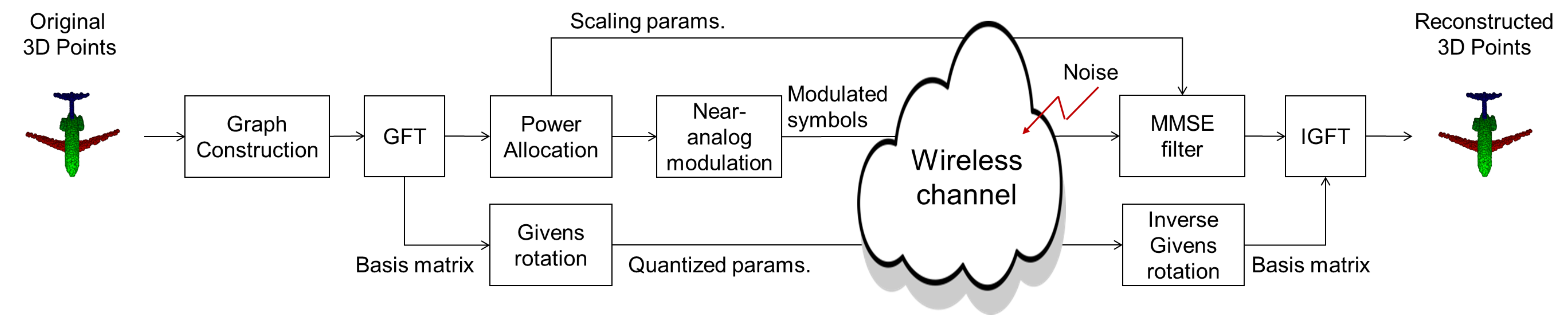}} \\
  \subfloat[Proposed GNN-Based Scheme]
  {\includegraphics[width=\linewidth]{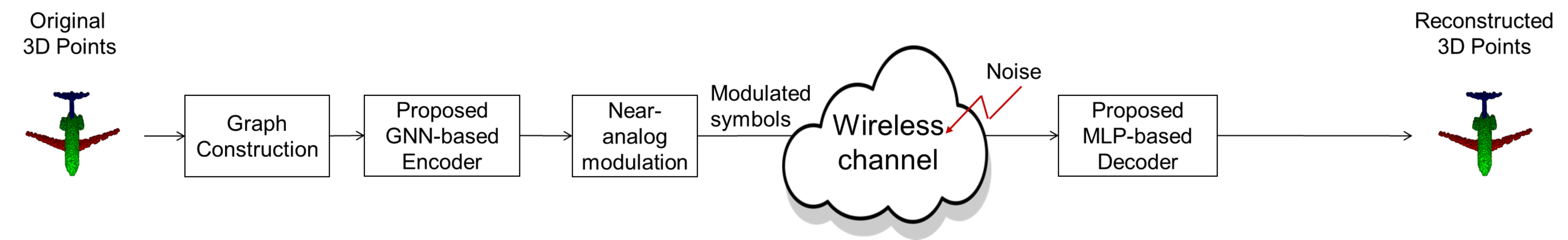}}
 \caption[]{Overview of conventional and proposed schemes for wireless 3D point cloud delivery.} 
 \label{fig:scheme}
 \end{figure*}

However, the conventional schemes of point cloud delivery suffer from the following problems due to the
wireless channel unreliability and nonlinear operations of data compression.  First, the encoded bitstream is highly
vulnerable for bit errors~\cite{bib:survey}.  When the channel signal-to-noise ratio (SNR) falls under a certain threshold, possible catastrophic errors occurred in the bitstream during communications will disable point cloud restoration.
This phenomenon is called the cliff effect~\cite{bib:fuji_GMRF}. 
Second, the reconstruction quality does not improve even when the wireless channel quality is improved unless an adaptive rate control of source and channel coding is performed in real-time according to the rapid fading channels. This is called the leveling effect.  
Third,
quantization is a lossy process and its distortion cannot be recovered
at the receiver. 
Finally, voxel-domain point cloud encoders~\cite{bib:PCL, bib:PCL2} have limited coding efficiency since it does not yield a good energy compaction. Although conventional transform techniques, such as discrete cosine transform (DCT), can be used even for point cloud data, they do not fully exploit the underlying irregular geometry of the 3D points.

To solve the above-mentioned issues, we have proposed HoloCast~\cite{bib:HoloCast, bib:HoloCastGivens} to realize graceful 3D reconstruction quality improvement with the improvement of wireless channel quality. Fig.~\ref{fig:scheme}(a) shows the overview of HoloCast. 
The key ideas of HoloCast are 1) skipping digital operations, i.e., quantization and entropy/channel coding, analogous to SoftCast~\cite{bib:SoftCast_second} and 2) introducing graph signal processing (GSP)~\cite{bib:GSP} to achieve better energy compaction.
Specifically, HoloCast regards the 3D points as vertices in a graph and takes graph Fourier transform (GFT)~\cite{bib:DigitalGFT} to exploit the correlations between the adjacent graph signals, and directly sends the GFT coefficients by using near-analog modulation~\cite{bib:SoftCast_second}.
However, the GFT-based coding in HoloCast needs a large communication overhead for decoding even with overhead reduction techniques~\cite{bib:HoloCastGivens}. 
Specifically, a sender needs to transmit the eigenvectors of graph Laplacian matrix as metadata.

The main objective of our study is to extend HoloCast by introducing a new framework known as graph neural networks (GNN)~\cite{bib:GNN} to realize high quality and low overheads. 
GNN is a novel model for graph representation learning, which allows analyzing irregular geometric structure of graph data. 
We focus on an end-to-end (E2E) deep learning, i.e., GNN-based autoencoders (GAE)~\cite{bib:GNNAE2017,bib:FoldingNet, bib:ChenDYLFT:20,bib:PCT}, to encode 3D point clouds into a compressed representation. 
One of the benefits in the GAE is to allow the graph signal reconstruction from the limited number of latent variables without requiring additional metadata. 
Fig.~\ref{fig:scheme}(b) shows the overview of the proposed scheme, where the GNN-based encoder compresses 3D points into the latent variables, and then the compressed variables are directly mapped to transmission signals without relying on digital modulation schemes.
The latent variables, which are distorted through wireless fading channels, are fed into another neural network decoder to reconstruct clean 3D points. 

Our contribution is three-fold:
1) we verify that the proposed GNN-based point cloud delivery realizes better 3D reconstruction quality compared with the conventional HoloCast over fading channels, 2) we confirm that the proposed GNN-based encoder can reduce the amount of communication overhead by one order of magnitude, and 3) we demonstrate that adaptive channel precoding brings further quality improvement by means of the diversity gain of the rapid fading channels.

\section{GNN-Based Soft Point Cloud Delivery}

\begin{figure*}[t]
  \begin{center}
   \includegraphics[width=\hsize]{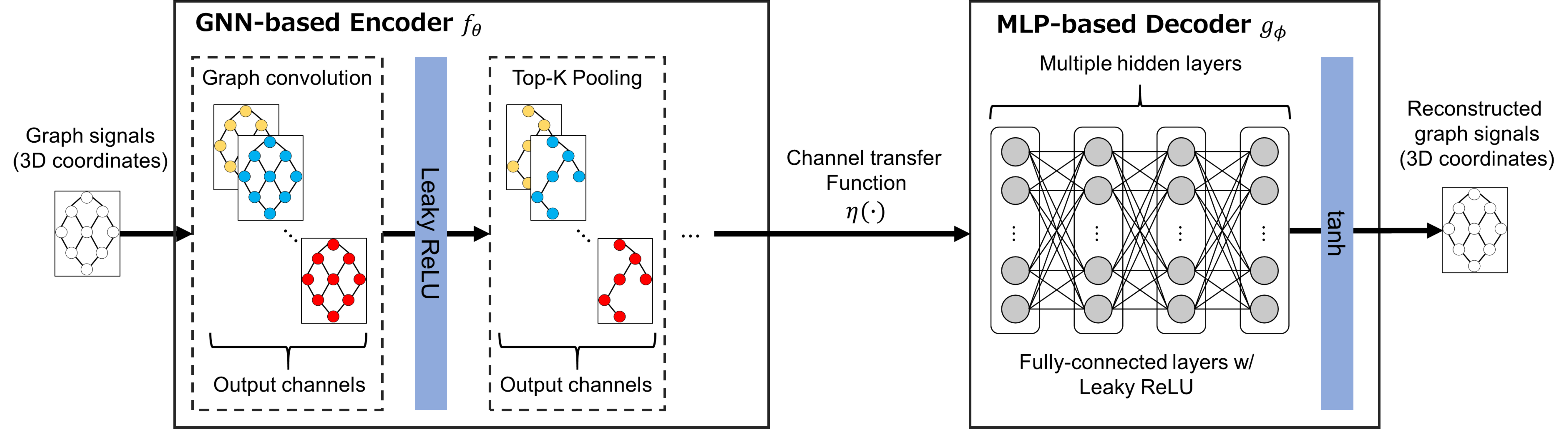}
   \caption{Proposed GNN-based end-to-end encoder and decoder for wireless 3D point cloud delivery.}
   \label{fig:NN}
  \end{center}
\end{figure*}

Fig.~\ref{fig:NN} shows the proposed E2E point cloud encoder and decoder, to prevent the cliff/leveling effects in 3D scene reconstruction, to gracefully improve reconstruction quality along with channel quality, and to reduce the amount of overhead.

\textbf{Encoder:} The encoder part first regards the 3D points as a graph signal using a weighted and undirected graph $\mathcal{G} = (\boldsymbol{V}, \mathcal{E}, \boldsymbol{W})$ where $\boldsymbol{V}$ and $\mathcal{E}$ are the vertex and edge sets of $\mathcal{G}$, respectively. $\boldsymbol{W}$ is an adjacency matrix having positive edge weights and the $(i,j)$th entry $\boldsymbol{W}_{i, j}$ represents the weight of an edge connecting vertices $i$ and $j$. 
We consider the attributes of the point cloud, i.e., the 3D coordinates $\bm{p} = [\mathsf{x}, \mathsf{y}, \mathsf{z}]^\mathrm{T} \in \mathbb{R}^{3 \times N}$ as signals that reside on the vertices in the graph ($N$ is the number of vertices). 
For simplicity, we consider a $K$-nearest-neighbor graph, where each 3D point connects to its $K$ closest 3D points. Here, we use a binary adjacency matrix whose entry is either is 1 or 0 to indicate connectivity. 
The encoder maps the 3D coordinate attributes $\bm{p}$ to $m$-dimensional and $L$-channel real-valued latent variables $\bm{z} \in \mathbb{R}^{m \times L}$ by means of an encoding function $f_\theta$. The encoding function $f_\theta$ is parameterized using graph convolutional neural networks (GCNN) with weights $\theta$.
The encoder consists of a series of graph convolution followed by leaky rectified linear unit (ReLU) activation function, Top-K pooling~\cite{bib:TopK}, and a normalization layer. The graph convolution layers extract the graph signal features and the nonlinear activation function allows to learn a non-linear mapping from the source signal to the coded signal. 
Top-$K$ pooling layer chooses the largest $K$ values from each channel to remain important features. The output of the last graph convolution layer is normalized such that
$
\|{\bm{z}}\|^2 = m L P$,
where $P$ denotes the average transmission power.

\textbf{Wireless Link:}  The coded variables $\bm{z}$ are sent over the communication channel by directly mapping to in-phase and quadrature (I-Q) symbols $\bm{x}$ for analog wireless transmissions. 
The wireless channel, denoted by $\eta$, introduces stochastic distortion to the transmission symbols. 
To optimize the proposed scheme under wireless communications, the channel transfer function $\eta$ must be incorporated into the E2E GAE. 
We consider the channel model based on Rayleigh fading as reasonable wireless communication systems.
In Rayleigh fading channels, each analog-modulated symbol at the receiver can be modeled as follows:
$y_i = h_i x_i + n_i$,
where $y_i$ is the $i$th received symbol, $x_i$ is the $i$th transmission symbol, $h_i$ is $i$th multiplicative fading coefficient, and $n_i$ is an additive white Gaussian noise (AWGN) with an average noise variance of $\sigma^2$. The fading coefficients in the Rayleigh fading channels follow zero-mean complex Gaussian distribution, i.e., $h_i \sim \mathcal{CN}(0,1)$ where $\sim$ means ``distributed as'' and $\mathcal{CN}(a,b)$ is a complex Gaussian distribution with a mean of $a$ and variance of $b$.

To reduce the impact of fading effects, we consider two equalization techniques at the sender and receiver, i.e., pre-equalization and post-equalization, for a channel transfer function $\eta$. 
The pre-equalization can be realized at the sender side by sending pre-equalized transmission symbol $x_i$ to the receiver as $x_i = w_i z_i$ where $w_i$ is a pre-equalizer weight.  Although there are many variants of pre-equalizer, we assume a simple pre-equalization: $w_i = h_i^* / |h_i|$ where $[\cdot]^*$ denotes the conjugate operation. In this case, the channel transfer function will be: $\eta_\mathrm{preeq}(z_i) = |h_i| {z}_i + n_i$.
The post-equalization can be realized at the receiver side by taking an inverse operation of the fading attenuation. Specifically, the receiver takes the post-equalization such that $\widehat{y}_i = y_i / h_i$, given the estimated fading coefficient $h_i$. In this case, the channel transfer function will be: $\eta_\mathrm{posteq}(z_i) = z_i + n_i / h_i$. 
In addition to pre-/post-equalization, we also consider precoding method which sorts the latent variables $\bm{h}$ according to the fading level $|h_i|$ in descending order. Such sorting may facilitate for GNN to optimize the best latent variables to achieve diversity gain. 

\textbf{Decoder:} Upon the receipt of distorted latent variables, the decoder uses a decoding function $g_\phi$, based on a multi-layer perceptron (MLP) for 3D point cloud reconstruction. The decoder consists of a series of fully-connected layers and leaky ReLU with a parameter set $\phi$. The MLP decoder maps the distorted latent variables $\bm{\widetilde{z}}$ into an estimate $\bm{\widehat{p}}$ of the 3D coordinates. The last layer uses hyperbolic tangent (tanh) activation function.

\textbf{Loss Function:} The proposed GNN-based encoding and decoding functions are trained to minimize a loss function:
\begin{equation*}
(\theta, \phi) = \argmin_{\theta, \phi} \mathop{\mathbb{E}}_{\Pr(\bm{p}, \hat{\bm{p}})} \big[d(\bm{p}, \hat{\bm{p}}) \big],
\end{equation*}
where $\mathbb{E}[\cdot]$ is an expectation, $d(\bm{p}, \hat{\bm{p}})$ is a defined distortion function between the original and reconstructed 3D coordinate attributes, $\Pr(\bm{p}, \hat{\bm{p}})$ is the joint probability distribution of the original and reconstructed 3D coordinate attributes. Since the true distribution of the input attributes is often unknown and thus the expected distortion is also unknown. 
To learn better weights for the minimization of the expected distortion in Rayleigh fading channels, all potential distortions due to channel fading and additive noise are synthetically analyzed by the proposed scheme in off-line learning phase.
We use adaptive momentum (ADAM) optimizer for weight learning with an initial learning rate of $0.005$, batch size of $10$, momentum of $0.9$, and momentum2 of $0.999$ for $500$ epochs.

\section{Performance Evaluation}
\label{sec:evaluation}
\subsection{Simulation Settings}
\noindent \textbf{Datasets:} We use a benchmark dataset of ShapeNet~\cite{bib:shapenet} for experiments.  ShapeNet contains more than $50{,}000$ unique 3D points from $55$ categories. In our experiments, we select point clouds of ``Airplane'' category. We sample $2{,}115$ point clouds for training and $234$ point clouds for testing.  The training data are used for learning the network weights while the testing data are used for comparison in terms of 3D reconstruction and visual quality.

\noindent \textbf{Quality Metric:} We use the augmented Chamfer distance~\cite{bib:ChenDYLFT:20} as the distortion function for the 3D coordinate attributes.
The augmented Chamfer distance $d_\mathsf{CH}(S, \widehat{S})$ is defined as
\begin{align*}
\max \bigg\{ \frac{1}{|S|} \sum_{\bm{p}\in S} \min_{\widehat{\bm{p}} \in \widehat{S}} \|\bm{p} - \widehat{\bm{p}}\|_2,   \frac{1}{|\widehat{S}|} \sum_{\bm{\widehat{p}}\in \widehat{S}} \min_{\bm{p} \in S } \|\bm{p} - \widehat{\bm{p}}\|_2 \bigg\},
\end{align*}
where $S$ is the input point set and $\widehat{S}$ is the reconstructed point set. The term $\min_{\widehat{\bm{p}} \in \widehat{S}} ||\bm{p} - \widehat{\bm{p}}||_2$ enforces that any 3D coodinate $\bm{p}$ in the original point cloud has a matching 3D point $\widehat{\bm{p}}$ in the reconstructed point cloud, and the term $\min_{\bm{p} \in S } ||\bm{p} - \widehat{\bm{p}}||_2$ enforces the matching vice versa. The max operation enforces that the distance from $S$ to $\widehat{S}$ and the distance vice versa have to be small simultaneously.

\noindent \textbf{Wireless Environment:}
We consider Rayleigh fading channels with an additive noise $n_i$ for realistic wireless environments. 
The additive noise $n_i$ follows white Gaussian distribution with a variance of $\sigma^2$, i.e., $n_i \sim \mathcal{CN}(0, \sigma^2)$.  

\noindent \textbf{GAE Architecture:} We use PyTorch Geometric (PyG)~\cite{bib:PyG} for the implementation of our GAE architecture. 
The encoder repeats a series of GCNConv~\cite{bib:GCN} with the output channels between $12$ and $48$, leaky ReLU activation function, and Top-K pooling at the graph pooling ratio between $0.5$ and $0.9$ three times. The output of the last Top-K pooling layer is followed by normalization layer which enforces the average power constraint. The decoder uses a series of fully-connected layer and leaky ReLU three times to reconstruct the 3D coordinate attributes from the distorted latent variables via a channel transfer function. Here, the output channels of the first and the second fully-connected layers are the same as the output channels of GCNConv while the output channels of the last fully-connected layer is $3$. 

\begin{figure}[t]
  \begin{center}
   \includegraphics[width=\hsize]{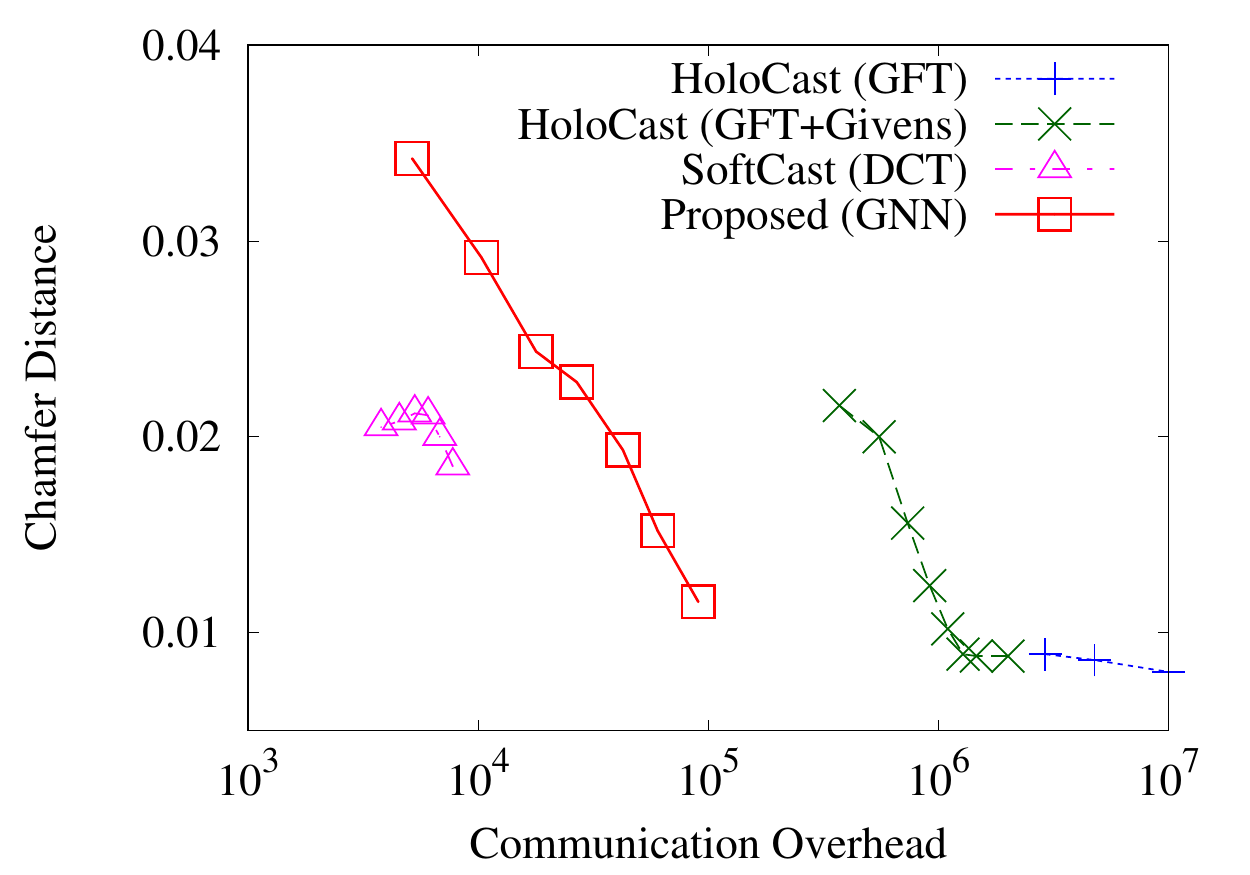}
   \caption{3D reconstruction quality as a function of communication overheads at a wireless SNR of $20$~dB. Here, we ignore the communication overhead for power scaling in the conventional SoftCast and HoloCast schemes.}
   \label{fig:metadistance}
  \end{center}
\end{figure}

\subsection{Overhead Reduction}
We firstly discuss an impact of the proposed GNN-based coding on the amount of communication overheads. 
Fig.~\ref{fig:metadistance} shows the 3D reconstruction quality over Rayleigh fading channels as a function of communication overheads at a wireless channel SNR of $20$~dB. Here, the communication overhead is the total number of transmission symbols in near-analog modulation (and GFT basis matrix for HoloCast). 

We compare four schemes: HoloCast~\cite{bib:HoloCast}, HoloCast with Givens rotation~\cite{bib:HoloCastGivens}, SoftCast~\cite{bib:SoftCast_second}, and proposed schemes. HoloCast uses octree decomposition and takes GFT for the graph signals in each octree block to convert into spectrum domain by using the eigenvectors of the random-walk graph Laplacian matrix. HoloCast sends the eigenvectors of varying octree block sizes without compression. HoloCast with Givens rotation uses a quantization bit depth $b$ between $2$ and $12$ to compress the eigenvectors for overhead reduction. Here, the octree block size is fixed to $300$. SoftCast takes DCT-based decorrelation for 3D coordinate attributes and directly maps the coefficients on the I-Q components. The proposed scheme uses GNN-based encoding and decoding for overhead reduction. Here, the proposed scheme uses precoding with a channel transfer function of post-equalization.

We can see that the proposed scheme achieves a significant overhead reduction at the same 3D reconstruction quality compared with the conventional HoloCast and HoloCast with Givens rotation. For example, Chamfer distance of the proposed scheme is $0.011$ at the communication overhead of $9.0 \times 10^4$ symbols. On the other hand, Chamfer distance of HoloCast with Givens rotation is $0.010$ at the communication overhead of $1.1 \times 10^6$ symbols. In this case, the proposed scheme achieves $92.0$\% overhead reduction at almost the same 3D reconstruction quality. 
The conventional SoftCast has a limited 3D reconstruction quality irrespective of the communication overhead.
Although the communication overhead in SoftCast is smaller than that in the proposed scheme, SoftCast needs an additional communication overhead for power scaling~\cite{bib:fuji_GMRF}. Detailed analysis of the communication overhead for power scaling will be left as the future work. 

\begin{figure}[t]
  \begin{center}
   \includegraphics[width=\hsize]{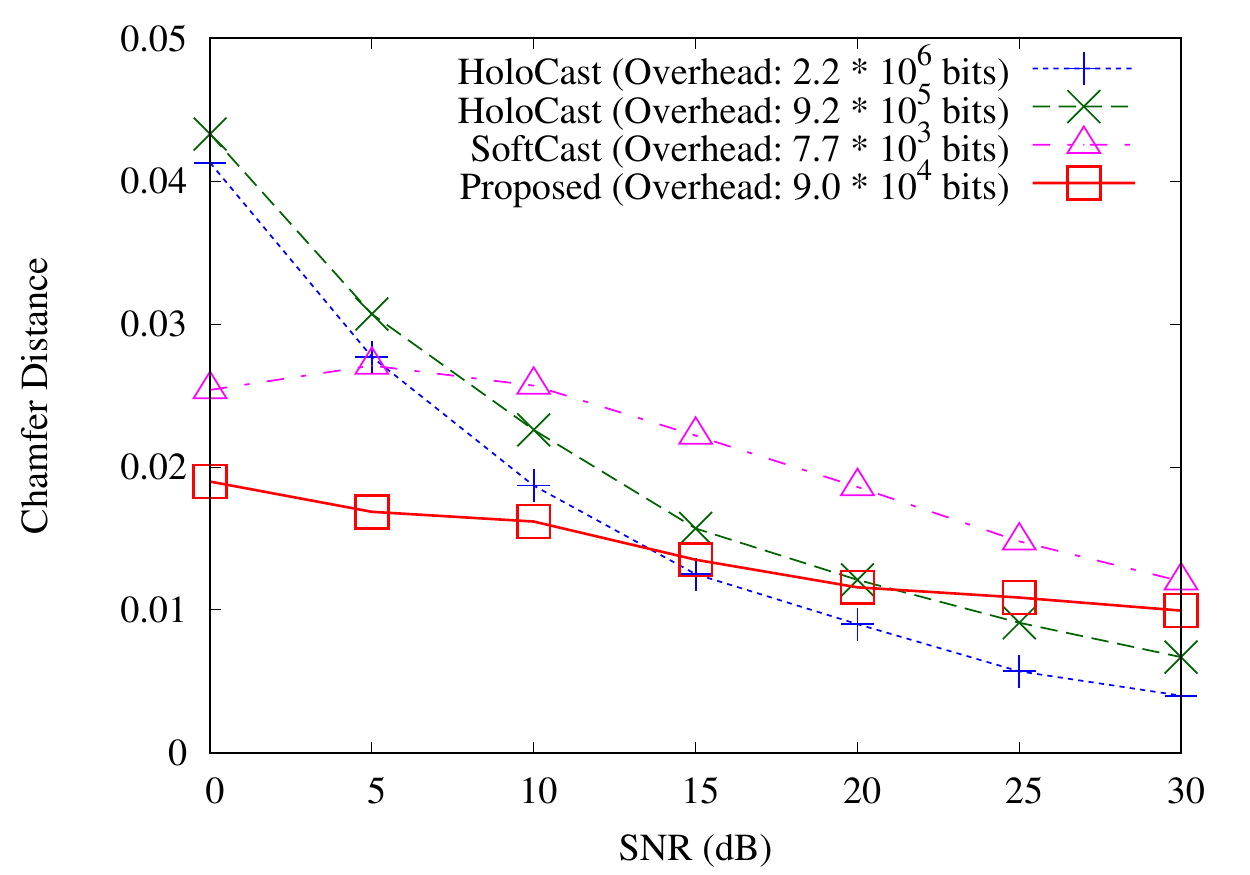}
   \caption{3D reconstruction quality as a function of wireless channel SNRs in Rayleigh fading channels. Note that the conventional HoloCast and SoftCast schemes ignore  the communication overhead for power scaling.}
   \label{fig:sndistance}
  \end{center}
\end{figure}

\subsection{3D Reconstruction Quality}

We now discuss an effect of wireless channel quality on the reconstructed point cloud quality.  We consider two HoloCast schemes with a bit depth of $5$ and $12$ in Givens rotation at an octree decomposition size of $300$. 

Fig.~\ref{fig:sndistance} shows the 3D reconstruction quality of the reference schemes over Rayleigh fading channels as a function of wireless channel SNRs. 
We observe the following results:
\begin{itemize}
  \item The proposed scheme yields the best 3D reconstruction quality in low wireless SNR regimes.  
  \item Although the conventional HoloCast schemes realize better 3D reconstruction quality in high wireless SNR regimes, the required overhead is more than $10$-times larger than that of the proposed method.  
  \item The 3D reconstruction quality of SoftCast is lower than that of the proposed scheme irrespective of wireless channel SNRs. 
  \end{itemize}

 \begin{figure}[t]
  \begin{center}
   \includegraphics[width=\hsize]{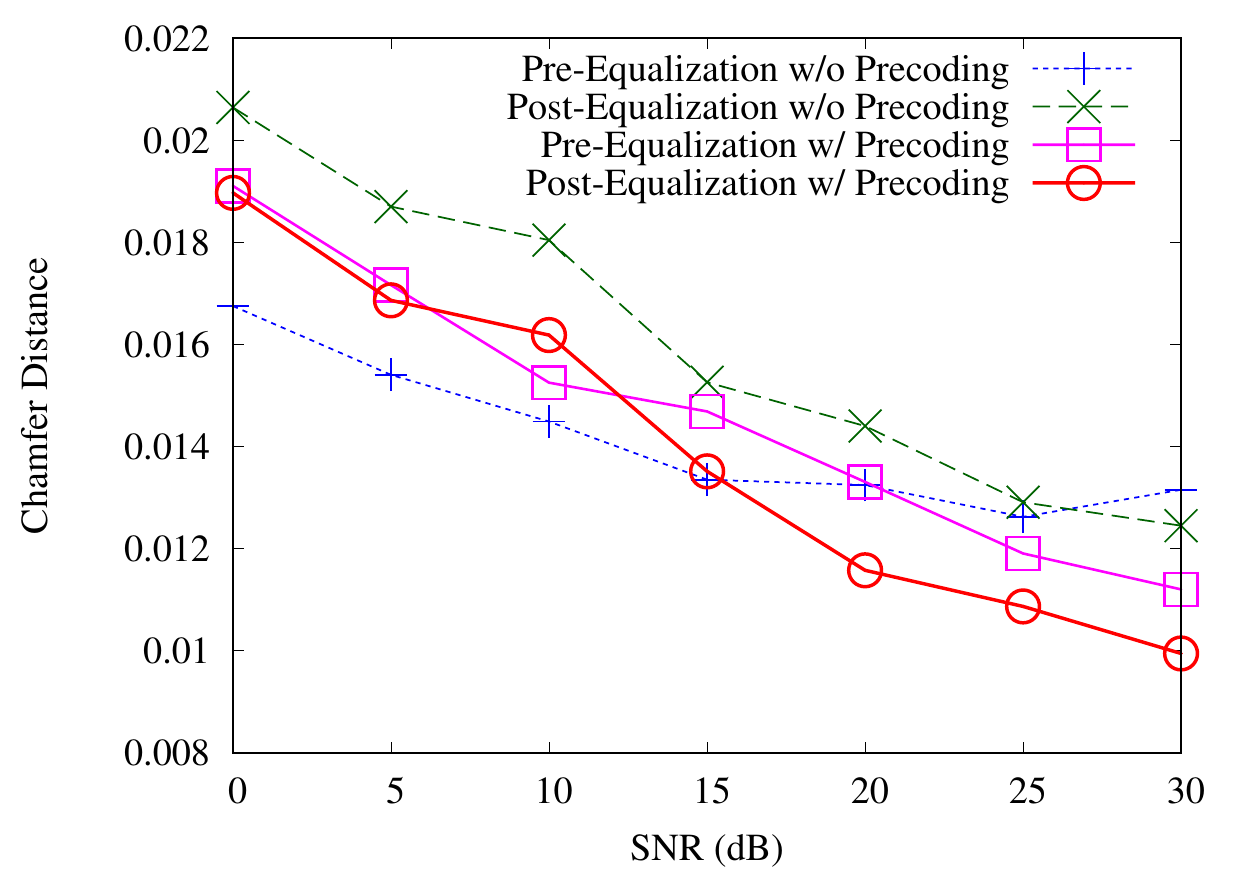}
   \caption{3D reconstruction quality of the proposed schemes over different channel transfer functions with/without precoding.}
   \label{fig:precoding}
  \end{center}
\end{figure}

\begin{figure}[t]
\centering
 \subfloat[Original]
 {\includegraphics[width=0.46\hsize, trim=25 50 25 50, clip]{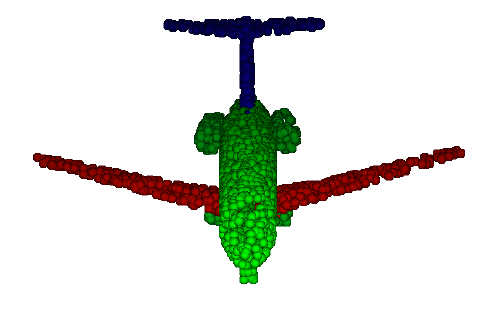} \label{fig:orig}}
 \hfill
 \subfloat[SoftCast (DCT) \newline Chamfer distance: $0.015$]
 {\includegraphics[width=0.48\hsize, trim=25 50 25 50, clip]{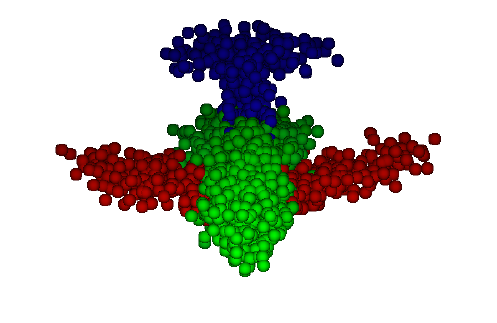} \label{fig:softcast}}
 \\
  \subfloat[HoloCast (GFT) \newline Chamfer distance: $0.013$]
 {\includegraphics[width=0.48\hsize, trim=25 50 25 50, clip]{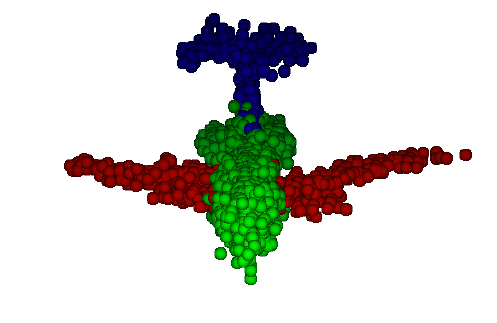} \label{fig:holocast}}
 \subfloat[Pre-equalization w/o Precoding \newline Chamfer distance: $0.013$]
 {\includegraphics[width=0.48\hsize, trim=25 50 25 50, clip]{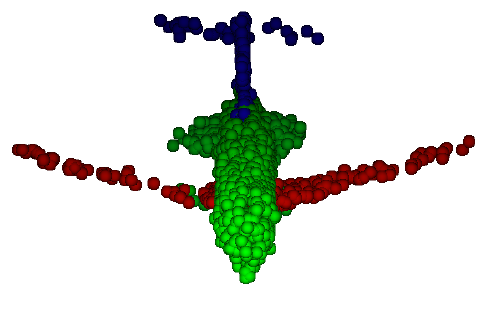} \label{fig:nopc_preeq}}
 \\
 \subfloat[Post-equalization w/o Precoding\newline Chamfer distance: $0.014$]
 {\includegraphics[width=0.46\hsize, trim=25 50 25 50, clip]{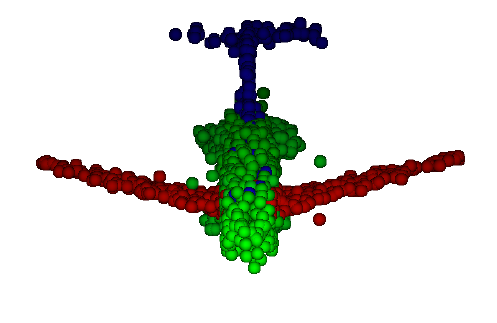} \label{fig:nopc_posteq}}
 \hfill
 \subfloat[Pre-equalization w/ Precoding \newline Chamfer distance: $0.013$]
 {\includegraphics[width=0.46\hsize, trim=25 50 25 50, clip]{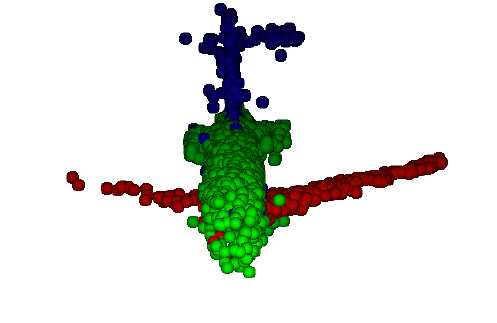}
 \label{fig:pc_preeq}}
 \\
 \subfloat[Post-equalization w/ Precoding \newline Chamfer distance: $0.012$]
 {\includegraphics[width=0.46\hsize, trim=25 50 25 50, clip]{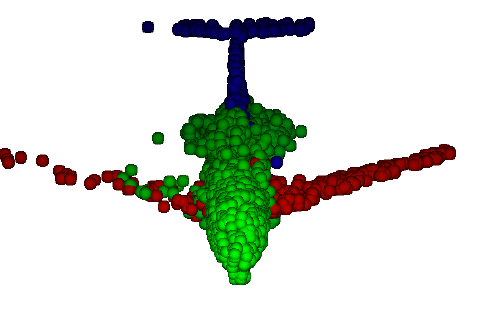}
 \label{fig:pc_posteq}}
 \caption{Snapshot of reconstructed 3D point cloud over different channel transfer functions with/without precoding at a channel SNR of $20$~dB.}
 \label{fig:snapshots}
\end{figure}

\subsection{Impact of Precoding and Equalization}
\label{sec:ev_precoding}

In this section, we evaluate the effects of precoding and channel transfer functions, i.e., post-equalization and pre-equalization, on the 3D reconstruction quality of the proposed scheme.
Fig.~\ref{fig:precoding} shows the 3D reconstruction quality of the proposed schemes over Rayleigh fading channels as a function of wireless channel SNRs for the case with different channel transfer functions of $\eta_\mathrm{preeq}$ and $\eta_\mathrm{posteq}$ with/without precoding. 
The evaluation results are summarized as follows:
\begin{itemize}
  \item Precoding performs well in high wireless SNR regimes since it may achieve a higher diversity gain.  
  \item Pre-equalization works well at lower SNR regimes, whereas post-equalization does well at higher SNRs.
  \item As a consequence, pre-equalization without precoding yields the best 3D reconstruction quality at low SNR regimes below $10$~dB.
  \item Accordingly, the post-equalization with precoding becomes the best one in the high SNR regimes above $10$~dB. 
 \end{itemize}

We finally compare some examples of visual snapshots for SoftCast, HoloCast, and the proposed schemes over Rayleigh fading channels in
Figs.~\ref{fig:snapshots}(a) through (g) at a channel SNR of $20$~dB. Here, the point cloud is selected from one point cloud from the test data in ShapeNet database. 
Although each proposed scheme may reconstruct the 3D shape of the aircraft, the proposed scheme with precoding may realize clear reconstruction compared with the proposed scheme without precoding. In particular, SoftCast has an obvious degradation over other schemes. Nevertheless, the 3D shape of the aircraft tail still remains noisy even with proposed methods. 
Note that we focused on a simplified GNN method compared to state-of-the-art techniques such as graph inception networks (GIN)~\cite{bib:ChenDYLFT:20} and FoldingNet~\cite{bib:FoldingNet} in order to demonstrate an initial proof-of-concept study of GNN-based 3D point cloud delivery.
Extension to further improve 3D reconstruction quality will be considered as another follow-up work. 
To the best of authors' knowledge, this paper is the very first study exploiting GNN methods for wireless 3D point cloud delivery.

\section{Conclusions}
\label{sec:conclusion}
We have proposed a novel scheme of soft point cloud delivery for future wireless streaming of holographic and 3D data.  
Specifically, the proposed scheme integrates GNN-based point cloud coding and near-analog modulation to simultaneously achieve: 1) prevention of the cliff effect, 2) prevention of the leveling effect, 3) high energy compaction, and 4) low communication overhead.
In addition, the proposed E2E design of GAE scheme accounts for random distortion due to fading channels through the use of pre-/post-equalization and precoding techniques. 
Evaluation results demonstrated that the proposed scheme achieves a good trade-off between 3D reconstruction quality and communication overhead compared with the conventional SoftCast and HoloCast. More rigorous analysis with GIN and FoldingNet will follow as future work.

\section*{Acknowledgment}
T. Fujihashi was partly supported by JSPS KAKENHI Grant Number 17K12672.

\bibliographystyle{IEEEtran}
\bibliography{hybrid}

\end{document}